\documentclass[amssymb,amsmath,aip,jcp,preprint]{revtex4-1}
\usepackage{bm}%
\usepackage[colorlinks=true,linkcolor=blue]{hyperref}%
\usepackage{graphicx}
\usepackage{epstopdf}
\usepackage{color,soul}
\soulregister\cite7
\soulregister\ref7
\soulregister\pageref7
\usepackage{subfig}
\usepackage{mathtools}
\usepackage{xr}
\expandafter\ifx\csname package@font\endcsname\relax\else
 \expandafter\expandafter
 \expandafter\usepackage
 \expandafter\expandafter
 \expandafter{\csname package@font\endcsname}%
\fi
\usepackage{graphicx}
\usepackage{epstopdf}
\usepackage{bm}
\usepackage[all]{xy}
\usepackage{afterpage}
\usepackage{amsmath}
\usepackage{float}
\usepackage{color}
\usepackage{enumerate}
\usepackage{enumitem, kantlipsum}
\expandafter\def\expandafter\normalsize\expandafter{%
    \normalsize
    \setlength\abovedisplayskip{40pt}
    \setlength\belowdisplayskip{40pt}
    \setlength\abovedisplayshortskip{40pt}
    \setlength\belowdisplayshortskip{40pt}
}

\begin{document}

\title{Two-dimensional Infrared Spectroscopy of vibrational polaritons of molecules in an optical cavity.}

\author{Prasoon \surname{Saurabh}$^a$}
\email{psaurabh@uci.edu}
\author {Shaul \surname{Mukamel}$^a$}
\email{smukamel@uci.edu}

\affiliation{$^a$Department of Chemistry, University
of California, Irvine, CA 92697}
\date{\today}




\begin{abstract}
Strong coupling of molecular vibrations to an infrared cavity mode affects their nature by creating dressed polariton states. We show how the single and double vibrational polariton manifolds may be controlled by varying the cavity coupling strength, and probed by a time domain 2DIR technique, Double Quantum Coherence (DQC). Applications are made to the amide-I ($CO$) and amide-II ($CN$) bond vibrations of $N-methylacetamide$ (NMA). 
\end{abstract}

\maketitle
\section{Introduction}

Elementary physical and chemical properties of molecules can be modified by coupling them to the optical modes of a cavity  thus forming strongly coupled $matter+field$ states known as polaritons \cite{berman1994cavity,haroche2006exploring}. These have been widely studied in atoms. Electronic polaritons in molecules have been extensively studied both experimentally and theoretically \cite{coles2014strong,coles2014polariton,herrera2014quantum}. Vibrational polaritons in the infrared has been recently demonstrated in molecular aggregates \cite{shalabney2015coherent,del2015quantum,shalabney2015enhanced,simpkins2015spanning} and in semiconductor nano-structures \cite{autry2015measurement,sun2015polaritons,wilmer2015multidimensional}. The radiation matter coupling is related to cavity frequency $\omega_c$ by,
$
 g_i =  \sqrt{N}\bm{\mu}_i\cdot \bm{e}_c \sqrt{\left( \hbar\omega_c / 2 \epsilon_0 V\right)},
$
 where, $N$ is the number of molecules. $\bm{\mu}_i$ is the transition dipole moment of the mode $i$, $\bm{e}_c$ is the cavity electric field vector, $\epsilon_0$ is vacuum permittivity, and $V$ is the cavity mode volume \cite{savona1995quantum, berman1994cavity,feist2015extraordinary,shalabney2015coherent,del2015quantum,shalabney2015enhanced,simpkins2015spanning}. While strong coupling to cavity modes has been realized even for a single atom, $N=1$,\cite{berman1994cavity,haroche2006exploring,faez2014coherent} polaritons in organic molecules were reported for large $N(\sim 10^{17})$ \cite{shalabney2015coherent}. With the advent of anomalous refractive index materials \cite{alu2007epsilon,fleury2013enhanced}, sub-wavelength Fabry-Perot microcavities \cite{kelkar2015sub}, and nanocavities \cite{benz2015control} it may become possible to achieve strong cavity coupling of single molecules.

Coherent multidimensional infrared spectroscopy is a powerful time domain tool that can probe anharmonicities and vibrational energy relaxation pathways \cite{lai2013monitoring,abramavicius2008double,venkatramani2002correlated,piryatinski2001two,greve2013n}. Vibrational polaritons were experimentally reported \cite{shalabney2015coherent,shalabney2015enhanced,simpkins2015spanning} and calculated \cite{del2015quantum,cwik2015self} recently. Multidimensional spectroscopic studies for electronically excited states in semiconductor cavity for nano-particles like $In_{0.04}Ga_{0.96}As$ has been demonstrated \cite{wilmer2015multidimensional,nardin2015multidimensional}. Similarly, electronic exciton-polariton interactions of quantum wells in microcavity \cite{takemura2015two} has also been reported. Bipolaritons generated using four wave mixing techniques have been used as efficient entangled photon source \cite{oka2008real,oka2008highly}. 

In this article, we calculate 2DIR signals for vibrational polaritons, focusing specially on the double quantum coherence (DQC) technique. Studying DQC of molecular vibrational polaritons in optical cavity can be used to study the effects of strong couplings on the vibrational anharmonicities and consequently allow us to control these anharmonicities. We introduce DQC signal in next section (Sec.\ref{dqci}), we then study single vibrational mode (Amide-I of NMA) coupled to a single mode cavity and calculate DQC in section (Sec.\ref{singlemode}), followed by two vibrational modes (Amide-I+II of NMA) coupled to a single mode cavity (Sec.\ref{tm}); and finally conclude in Sec.~\ref{CO}.

\section{Vibrational polaritons and their DQC signal} \label{dqci}
\begin{figure}[t]
\includegraphics[width=1\textwidth,bb=0 0 1156 646]{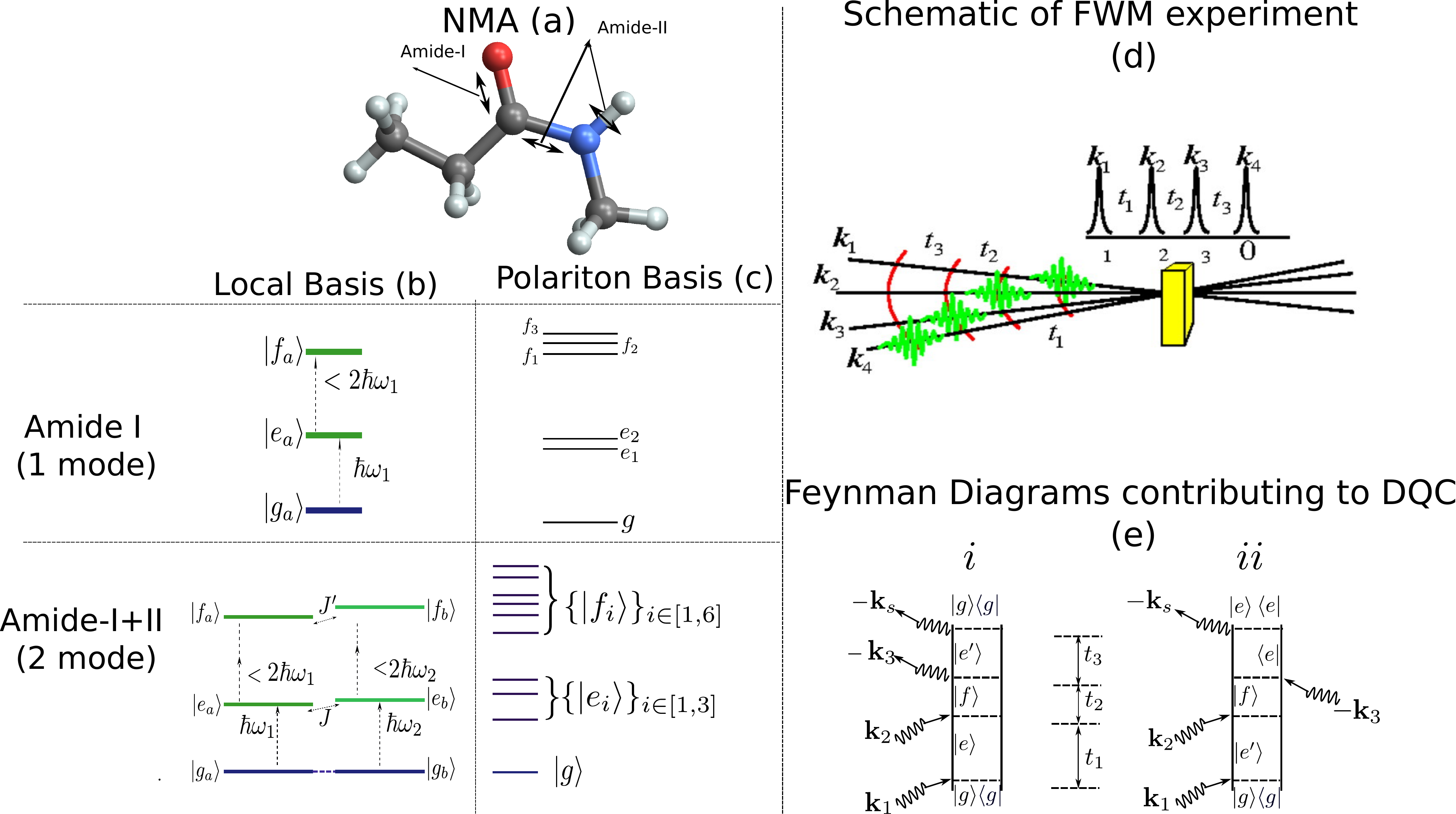}
\caption{Amide-I ($C=O$ symmetric stretch) and Amide-II ($CN$ symmetric stretch + $NH$ bend) vibrations are represented using double sided arrow for $N-Methylacetamide$ in (a). Eigen-energies of Amide-I and coupled Amide-I and Amide-II motifs in local basis and polariton basis (see Sec.SI of supplementary information \cite{supp} for detail) are given in (b) and (c) respectively. (d) schemes the FWM experiment, the yellow sample is the DBR Fabry-Perot cavity coupled to molecular vibrations (see text for detail) and, (e) shows the relevant ladder diagrams contributing to double quantum coherence (DQC) technique.\label{fig1a}}
\end{figure}

Vibrational modes (Fig.~\ref{fig1a}) coupled to an infrared cavity under the Rotating Wave Approximation (RWA) are described by the hamiltonian \cite{chernyak1998multidimensional,abramavicius2009coherent,roslyak2010two,shalabney2015coherent,del2015quantum,shalabney2015enhanced,simpkins2015spanning},
\begin{eqnarray}
\label{ex-cavham1}
H_0 &=& \omega_c(\theta) a^{\dagger} a + \sum_{i}^{m}\omega_{i}b_i^{\dagger} b_i +\sum_{i\neq j}^{m} J_{ij} b_i^{\dagger}b_j  \nonumber\\
&-&\sum_{ij}^{m} \frac{\Delta_{ij}}{2} b_{i}^{\dagger} b_{j}^{\dagger}b_{i} b_{j} +\sum_{i,j=1}^{m} g_{i} \left( a^{\dagger} b_{i}  +  b_{i}^{\dagger} a \right),
\end{eqnarray}
where, $a(a^{\dagger})$ and $b(b^{\dagger})$  are annihilation(creation) operators for the cavity photon, and vibrational excitation respectively, which satisfy boson commutation relations,  $[a,a^{\dagger}]=1;[b_i,b_j^{\dagger}]=\delta_{ij}$ and $\hbar=1$. $\omega_c(\theta) = \omega_0 \left(1-\frac{\sin^2(\theta)}{n^2_{eff}}\right)^{-1/2}$, is the angle-dependent cavity energy with  $\omega_0$ being cavity cut-off (or maximum) energy  and, $\theta$ the angle of incidence to the cavity mirrors. We set $\theta=0 ^{\circ}$ for simplicity. 
$\omega_i$ is vibrational frequency of mode $i$. $J_{ij}$ is the scalar coupling between two vibrational modes $i$ and $j$, while $\Delta_{ij}$ is the anharmonicity between respective modes. Finally, the coupling strength of vibrational modes $i$ to an optical mode is, $g_i$ as described in introduction.

The cavity volume $(V= (\lambda/n_{eff})^3)$ depends on cavity resonance wavelength ($\lambda$) and effective intra-cavity refractive index ($n_{eff}$) \cite{savona1995quantum, berman1994cavity,feist2015extraordinary}. Decreasing $n_{eff}$ can also be used for strong coupling to single molecular vibrational excitation, for instance, when the refractive indices of one of the two layers' of a Distributed Bragg Reflector (DBR) Fabry-Perot is equal to the empty cavity refractive index $n_c$, say $n_c=(n_2 \veebar n_1)$, then $n_{eff}=\sqrt{n_1n_2}$. In this case, choosing either one of the layers to be material with anomalous refractive index may decrease $n_{eff}<1$ \cite{kelkar2015sub}. The vacuum Rabi splitting $\Omega_R$ of mode $i$ is $2 \hbar g_i$.

\begin{figure}[t]
\includegraphics[width=0.55\textwidth,bb=0 0 1012 1285]{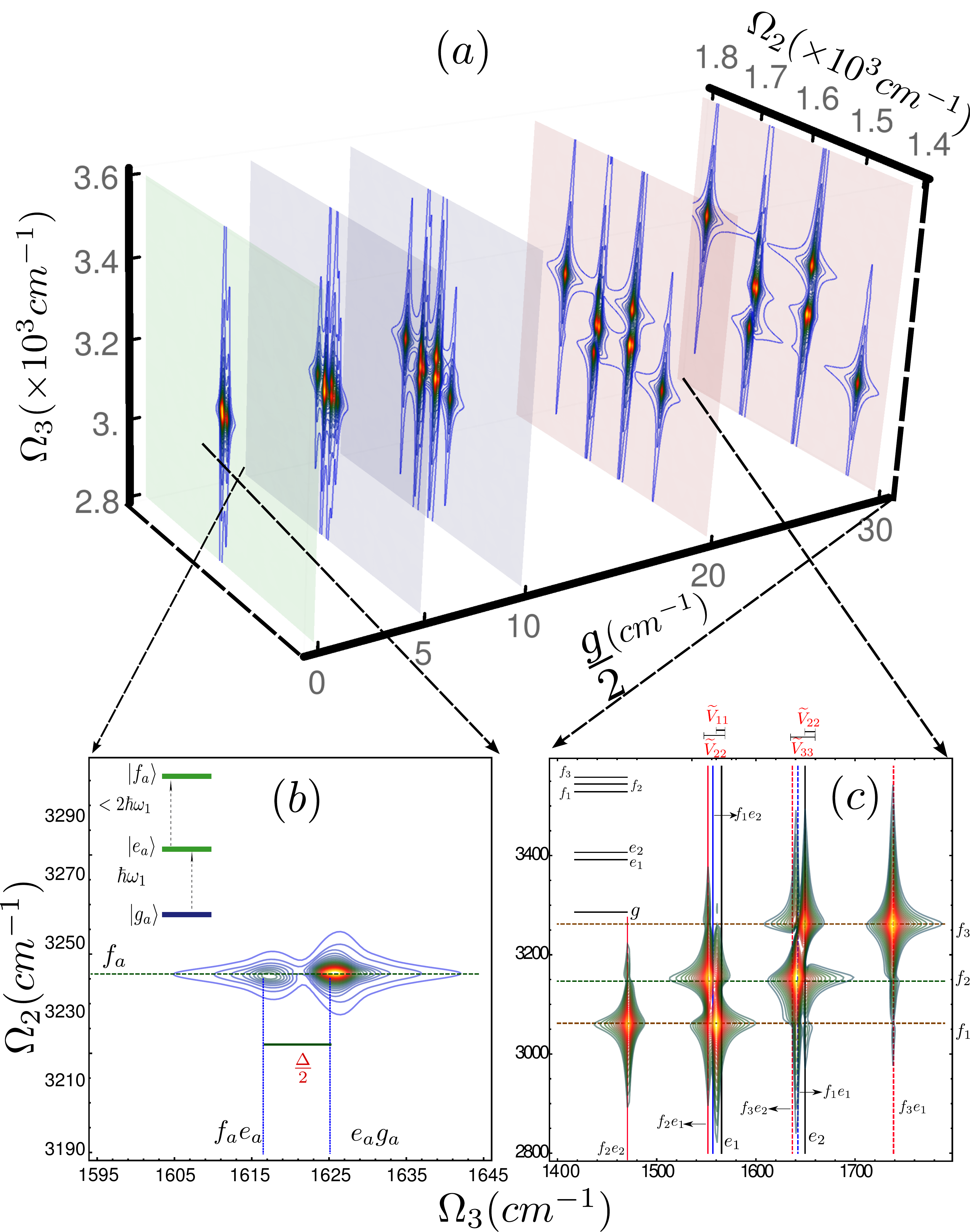}
\caption{(a) Absolute value of DQC signal (Eq.~\ref{Sigplot}) for Amide-I vibrations of NMA, as we vary coupling strength ($g$) from no (green plane), weak (blue planes) and strong (red planes) coupling regimes. (b) Schematic representation of DQC signal (zoomed in the grey square at $g=0 cm^{-1}$) with all possible peaks identified as intersections of different lines. (c) Same result but for strong coupling and the last panel. The lines are: i) $\Omega_{fe}$ labelled as $f_ie_j$ (dashed blue lines), ii) $\Omega_{eg}$ labelled as $e_j$ (black solid lines), iii) $\Omega_{fg}$ labelled as $f_i$ and appropriate anharmonicities ($\Delta_{ij}$ in local basis or $\widetilde{V}_{ii}$ in polariton basis) respectively. We wish to spectrally resolve (b) to (c).   \label{schematic}}
\end{figure}

DQC is a four wave mixing signal generated by three chronologically ordered pulses with wavevectors $\bm{k}_1$, $\bm{k}_2$ and $\bm{k}_3$ and detected by with a fourth pulse in the direction, $\bm{k_{III}}=\bm{k}_1+\bm{k}_{2}-\bm{k}_{3}$ (Fig.\ref{fig1a} (e)) \citep{abramavicius2008double,mukamel1999principles}. The signal is recorded versus three time delays $t_1$, $t_2$ and $t_3$. We assume a three polariton manifold as  shown in Fig.~\ref{fig1a}. 
Using the ladder diagrams for DQC (Fig.~\ref{fig1a}e) in polariton basis, which diagonalizes the first three and last terms of Eq.~\ref{ex-cavham1} (Fig.~\ref{fig1a}, Sec.SI of supplementary information \cite{supp}), we see that system oscillates with frequency $\Omega_{eg}=\Omega_e-\Omega_g$ during time delay $t_1=\tau_2-\tau_1$ and with frequency $\Omega_{fe}=\Omega_f-\Omega_e$ during delay $t_2=\tau_3-\tau_2$ in both contributing diagrams (Fig.~\ref{fig1a}e). After the third pulse the system oscillates either with frequency $\Omega_{e^{\prime}g}$ or $\Omega_{e^{\prime}f}$ during the delay $t_3=\tau_4-\tau_3$. For harmonic case, $\Omega_{e^{\prime}g}= \Omega_{e^{\prime}f}$ where the DQC signal vanishes. A 3D signal $\mathcal{S}(t_3,t_2,t_1)$, which can be written as double Fouier transform with respect to  $t_2$ and $t_3$ as \citep{abramavicius2008double}, 
\begin{eqnarray} \label{stimedomain}
\mathcal{S}(\Omega_3,\Omega_2,t_1)= \int \int_{0}^{\infty}dt_3dt_2 e^{i (\Omega_3 t_3+\Omega_2 t_2)} \mathcal{S}(t_3,t_2,t_1).  
\end{eqnarray} 
Upon expanding in polariton eigenstates, the signal with time delay $t_1=0$ becomes, 
\begin{eqnarray}
\label{Sigplot}
\mathcal{S}(\Omega_3,\Omega_2,t_1=0) &=&  \sum_{ee^{\prime},f} \frac{1}{(\Omega_2-\Omega_{fg}+i\gamma_{fg})}\nonumber\\
&& \bigg[\mu_{e^{\prime}f}\mu_{ge^{\prime}}\mu_{fg}\mu_{ge}
\frac{1}{\Omega_3-\Omega_{e^{\prime}g}+i\gamma_{e^{\prime}g}}\nonumber\\&-& \mu_{ge^{\prime}}\mu_{e^{\prime}f}\mu_{fg}\mu_{eg}
\frac{1}{\Omega_3-\Omega_{fe}+i\gamma_{fe}}\bigg].
\end{eqnarray}
Where, $\mu_{ij} $ is the polariton transition dipole from states in manifold $j \to i$ while $\gamma_{ij}$ is the respective dephasing. Note that the polariton eigenstates depend on the effective cavity coupling strengths, $\widetilde{g}_i \left(= \sqrt{Ng^{2}_i-\frac{1}{4}(\kappa-\gamma_i)^2}\right)$, where $\kappa$ is cavity decay rate, and $\gamma_i$ is dephasing rate of respective mode \cite{savona1995quantum}. Tuning the cavity coupling allows to control spectral structures of the singly and doubly excited vibrational polariton manifolds (Fig.\ref{schematic} and can be captured using DQC as illustrated in the following section (Sec.~\ref{rdm1}). 

\section{ A Single vibrational mode coupled to single cavity mode} \label{singlemode}\label{rdm1}
 The Amide motifs ($O=C-N-H$) link the amino-acid in peptides containing fundamental structural information, for instance, backbone geometry, interactions with hydrogen bonds and dipole-dipole interactions \cite{hayashi2008two}. 2D spectroscopy of the amide-I and II symmetric vibrations ( Fig.~\ref{fig1a}e ) have been studied \cite{rubtsov2003vibrational,lai2013monitoring}. We focus on the Amide-I vibrations in NMA (Table.~\ref{tb2}). We consider a single Amide-I stretch mode in resonance with the cavity mode and large anharmonicity $(\Delta_{ij} > \gamma_{ij}$) with varying coupling strengths $\widetilde{g}_1$ ranging from $0$ to $80$ $cm^{-1}$ (Table. \ref{tb2}). For a single vibrational mode (Fig.\ref{fig1a}), the three polariton manifolds have a ground state ($g$), two single excited states ($e$) and three doubly excited states ($f$). Furthermore, for detuning ($\delta = \omega_1-\omega_c$), the polariton basis modifies the anharmonicity $\Delta_{11} (\to \widetilde{V}_{11}=16\Delta_{11} |\widetilde{g}_1^4/(\delta^2-16 \widetilde{g}_1^2)^2 |)$.
\begin{table}
\caption{Parameters used in this work. $\omega_0$ is cavity cut-off frequency. $\omega_{1,2}$ are vibrational exciton (1,2) ($Amide-I,Amide-II$) energies \citep{barth2007infrared,deflores2006anharmonic}, $\gamma_{1,2}$ are their respective dephasing rates, $J_{ij}$ is harmonic scalar coupling, and $\Delta_{ij}$ are exciton-exciton interaction energies. \label{tb2}}
\centering
\begin{ruledtabular} 
\begin{tabular}{llcc}
(in $cm^{-1}$ )& Cavity & Amide-I & Amide-I+II\\
\hline            
Energy  & $\omega_0$=1625 & $\omega_1$=1625 & $\omega_1$=1625, $\omega_2$=1545\\
Dephasing  & $\kappa$=0 & $\gamma_1$=20 & $\gamma_1=\gamma_2$=20\\
\begin{minipage}[l]{.15\textwidth} 
Anharmonicities\\
(Local basis)
\end{minipage}  & $\Delta_{00}=0$ & $\Delta_{11}$=15 & \begin{minipage}{0.15\textwidth}$\Delta_{11}$=15,$\Delta_{22}$=11\\ $\Delta_{21}=\Delta_{12}$=10 \end{minipage}\\
Scalar coupling & - & - & $J_{12}$=15  \\
Eff. refractive ind. &$n_{eff}$ =0.5 & - & - \\
\begin{minipage}{.15\textwidth} 
Anharmonicities\\
(Polariton basis)
\end{minipage}  & \multicolumn{3}{c}{$\widetilde{V}_{ij}=\frac{\Delta_{ij}}{2}|X_i|^2|X_j|^2$}  
\end{tabular}
\label{table:nonlin}
\end{ruledtabular}
\end{table}

\begin{figure}[th]
\includegraphics[scale=0.3,bb=0 0 1442 569]{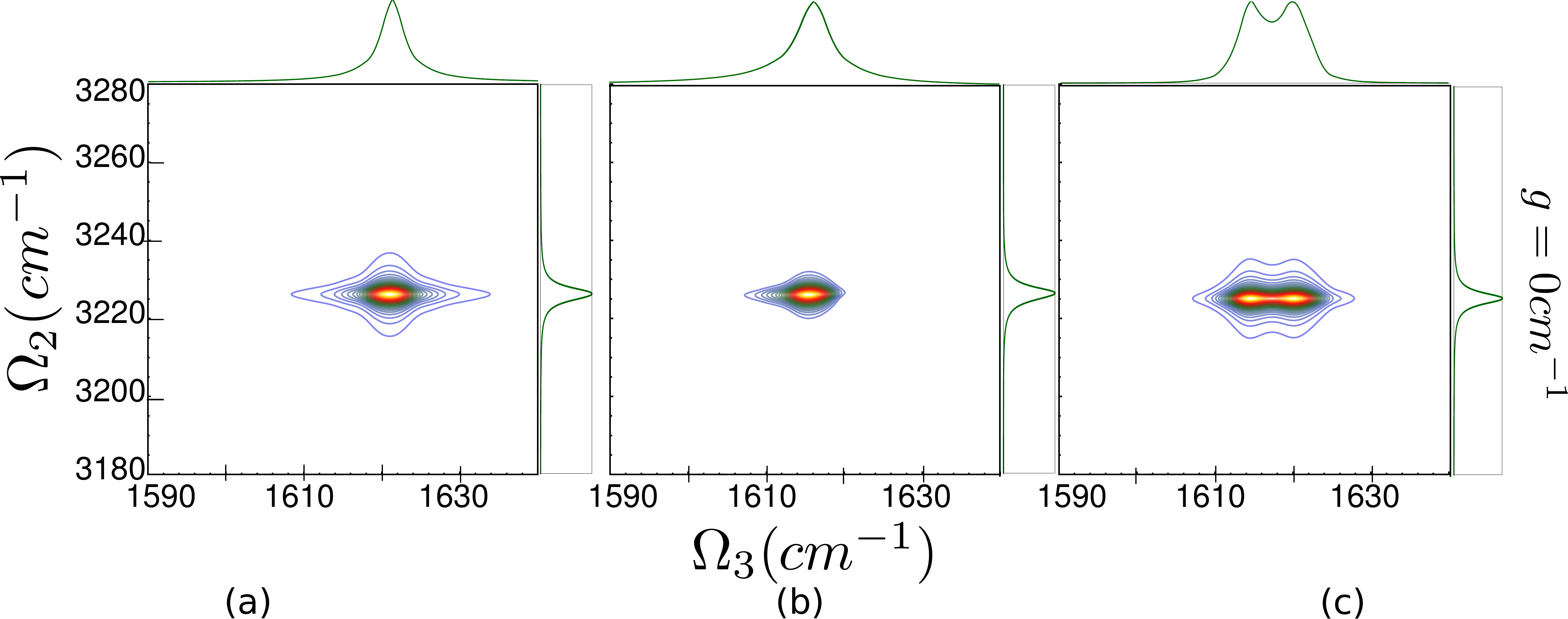}
\caption{DQC signals for Amide-I vibrations in NMA using Eq.~\ref{Sigplot} and ladder diagrams (Fig.~\ref{fig1a}e): (a) $|\mathcal{S}_i(\Omega_3,\Omega_2,0)|$ with resonance at $\Omega_{eg}=1625cm^{-1}$, (b)$|\mathcal{S}_{ii}(\Omega_3,\Omega_2,0)|$ with resonance at $\Omega_{fe}=1617 cm^{-1}$ and (c)$|\mathcal{S}(\Omega_3,\Omega_2,0)|$ with resonances at $\Omega_{eg}=1625cm^{-1}$ and $\Omega_{fe}=1617 cm^{-1}$. The linear projections along each axis is shown in green. For simplicity, cavity field vector $\bm{e}_c$ is assumed to be mostly parallel to molecular vibrational transition dipole $\mu_m$. Amide-I vibrations ($\omega_1$) is resonant to cavity cutoff frequency $\omega_0=1625 cm^{-1}$. The anharmonicity are assumed $\Delta_{11}=15cm^{-1}$ (Table ~\ref{tb2}).\label{nocav}}
\end{figure}

\begin{figure}[th]
\includegraphics[scale=0.30,bb=0 0 1725 1129]{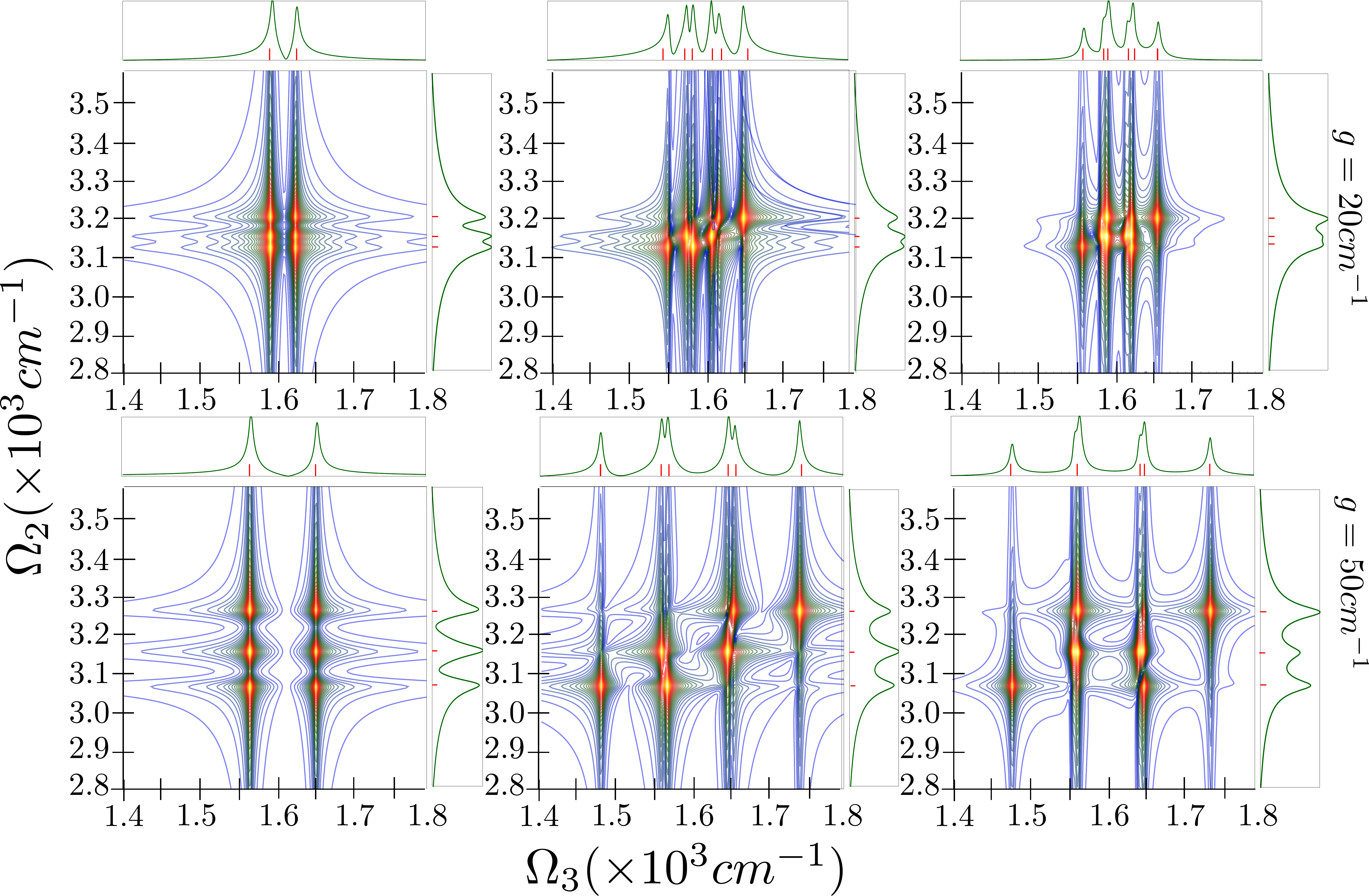}
\caption{DQC signals using Eq.~\ref{Sigplot} and ladder diagrams (Fig.~\ref{schematic}b): $Columns$ $ (\text{left}-)|\mathcal{S}_i(\Omega_3,\Omega_2,0)|$ with resonance at $\Omega_{e_1g}$ and $\Omega_{e_2g} $; $(\text{middle}-)|\mathcal{S}_{ii}(\Omega_3,\Omega_2,0)|$ with resonance at $\Omega_{f_2,e_2} $, $\Omega_{f_2,e_1} $, $\Omega_{f_1,e_2} $, $\Omega_{f_3,e_2} $, $\Omega_{f_1,e_1} $, $\Omega_{f_3,e_1} $ ; and $(\text{right}-)|\mathcal{S}(\Omega_3,\Omega_2,0)|$ with respective resonances for Amide-I vibrations in NMA with cavity coupling $rows:$ $(a) 20 cm^{-1}$ and $(a) 50 cm^{-1}$. The linear projections along each axis is shown in green. For simplicity, cavity field vector $\bm{e}_c$ is assumed to be mostly parallel to molecular vibrational transition dipole $\mu_m$. Amide-I vibrations ($\omega_1$) is resonant to cavity cutoff frequency $\omega_0=1625 cm^{-1}$. The anharmonicity are assumed $\Delta_{11}=15cm^{-1}$ (Table ~\ref{tb2}) \label{wgcav}}
\end{figure}

Using Eq.~\ref{Sigplot} and ladder diagrams (Fig.~\ref{fig1a}e) the signals $|\mathcal{S}_i(\Omega_3,\Omega_2,0)| $, $|\mathcal{S}_{ii}(\Omega_3,\Omega_2,0)| $ and $|\mathcal{S}(\Omega_3,\Omega_2,0)=|\mathcal{S}_{i}(\Omega_3,\Omega_2,0)+\mathcal{S}_{ii}(\Omega_3,\Omega_2,0)|$ are shown in Fig.~\ref{nocav} and \ref{wgcav} in left, center and right columns respectively (sec.~\ref{rdm1}). Absolute values are shown to illustrate peak assignments and affects of varying coupling strengths ($\widetilde{g}_{i}$). For completness, the respective absorptive (Im) and dispersive (Re) parts of the DQC signals are shown in Sec.SIIIA in \cite{supp}. The peak splittings depends on coupling strengths ($\widetilde{g}_i $) and polariton anharmonicities ($\widetilde{V}_{ij} $) and are given Sec.SV of ~\cite{supp}. 

\textit{Free molecule (Fig.~\ref{nocav}):} The Amide-I vibrations in NMA (without cavity) is a simple three level system as shown in Fig.~\ref{fig1a}. Under this condition, $e_1=e_2$ and $f_1=f_2=f_3$, thus we only observe single peak resonant at $\Omega_{fg}=3168cm^{-1}$ on $\Omega_2$ axis. The peaks due to $|\mathcal{S}_i(\Omega_3,\Omega_2,0)| $ (Fig.~\ref{nocav}(a)) shows $\Omega_{eg}=1625cm^{-1}$ resonance, while $|\mathcal{S}_{ii}(\Omega_3,\Omega_2,0)|$ (Fig.~\ref{nocav}(b)) shows resonance at $\Omega_{fe}=1617 cm^{-1}$  along $\Omega_3$ axis. The total signal (Fig.~\ref{nocav}(c)) thus shows two peaks due to resonances at  $\Omega_{eg}$ and $\Omega_{fe}$.

\textit{Weak coupling regime (Fig.~\ref{wgcav}-$\text{top row}$) with $\widetilde{g}$=$20cm^{-1}$:} Upon varying the coupling strength, there are two singly excited and three doubly excited polariton states (Fig. ~\ref{fig1a} b-c, Sec SII-S1 of \cite{supp} ). We thus observe that all $|\mathcal{S}_i(\Omega_3,\Omega_2,0)|$, $|\mathcal{S}_{ii}(\Omega_3,\Omega_2,0)| $ and $ |\mathcal{S}(\Omega_3,\Omega_2,0)|$ show three distinct peaks along $\Omega_2$ axis with energies resonant to $\Omega_{f_1g}=2\Omega_{e_1g}-\widetilde{V}_{11}$, $\Omega_{f_2g}=2\Omega_{e_1g}$ and $\Omega_{f_3g}=2\Omega_{e_2g}-\widetilde{V}_{33}$ respectively. Along $\Omega_3$ axis, $|\mathcal{S}_i(\Omega_3,\Omega_2,0)|$ (Fig.~\ref{wgcav}(top,left)) shows peak resonance at $\Omega_{e_1g}$ and $\Omega_{e_2g}$ with splitting $\sim \widetilde{g}$. The $|\mathcal{S}_{ii}(\Omega_3,\Omega_2,0)|$ (Fig.~\ref{wgcav}(top, middle)) has six distinct peaks at energies resonant to $\Omega_{f_2,e_2} $, $\Omega_{f_2,e_1} $, $\Omega_{f_1,e_2} $, $\Omega_{f_3,e_2} $, $\Omega_{f_1,e_1} $, $\Omega_{f_3,e_1} $ in increasing order respectively. The $|\mathcal{S}(\Omega_3,\Omega_2,0)|$ (Fig.~\ref{wgcav}(top, right)) has diminished peaks at $\Omega_{f_2e_1}$ and $\Omega_{f_3e_2}$ due to destructive interference of $\Omega_{f_2e_1} $ with $\Omega_{e_1g}$ and $\Omega_{f_3e_2} $ with $\Omega_{e_2g}$ resonances.

\textit{Strong coupling regime (Fig.~\ref{wgcav}-$\text{bottom row}$) with $\widetilde{g}=50cm^{-1}$: }
The signals corresponding to $|\mathcal{S}_{i}(\Omega_3,\Omega_2,0)|$ (Fig.~\ref{wgcav}(bottom, left)), $|\mathcal{S}_{ii}(\Omega_3,\Omega_2,0)|$ (Fig.~\ref{wgcav}(bottom, middle)) and $|\mathcal{S}(\Omega_3,\Omega_2,0)|$ (Fig.~\ref{wgcav}(bottom, right)) show peaks corresponding to $\Omega_{f_1g}$, $\Omega_{f_2g}$ and $\Omega_{f_3g}$ along $\Omega_2$ axis. Projections on $\Omega_3$ axis for $|\mathcal{S}_{ii}(\Omega_3,\Omega_2,0)| $ six peaks assigned as in weak coupling case, however the peak splitting between resonance pairs with energies $(\Omega_{f_2e_1},\Omega_{f_1e_2})$ and $(\Omega_{f_3e_2},\Omega_{f_1e_1})$ remains relatively constant and proportional to respective contributing anharmonicities. The total signal shows six peaks due to $\Omega_{f_2e_1}$ and $\Omega_{f_3e_2}$ diminished because of destructive interferences from resonances of singly excited polaritons.

This partial cancellation of peaks puts an upper bound on strong coupling strengths for fully resolved DQC signals. Furthermore by changing $t_1$, we can observe the dynamics in singly and doubly excited polariton manifolds to obtain bountiful information regarding lifetimes of molecular vibrational bipolaritons. In case of zero detuning ($\delta=\omega_1-\omega_c=0$), the doublet peak splitting due to $\widetilde{V}_{11}$ is independent of coupling strength $i.e.,\widetilde{V}_{11}=\Delta /32$. Thus DQC can be used as direct measurements of anharmoncities due to vibrational polariton-polariton interactions for vanishing detuning cases. One has to be careful however, for $\delta \neq 0$  cases; as such conditions allow the anharmonicities to vary non-trivially ($\sim 16\Delta_{11} |\widetilde{g}_1^4/(\delta^2-16 \widetilde{g}_1^2)^2 | $). This may cause destructive interference of different peaks along $\Omega_3$ axis.


\section{Polaritons for two vibrational modes and their DQC signal} 
\label{tm}

We now couple the Amide-I and Amide-II vibrations (Fig.~\ref{fig1a}, \cite{rubtsov2003vibrational}) of a single NMA molecule ($N=1$) to an infrared cavity. We next calculate the double quantum coherence (DQC) signals for vibrational molecular polaritons (Fig.~\ref{fig1a} (e), Eq.~\ref{Sigplot}) utilizing Eqs.S2-S5 (of supplementary material \cite{supp}).  

The overlapping of peaks due to anharmonicities is better illustrated in case of Amide-I+II vibrations coupled to single mode cavity. We next present three coupling regimes for such a condition. We assume $\widetilde{g}_1/\widetilde{g}_2=constant$ for simplicity. The remainder of relevant parameters are shown in Table ~\ref{tb2}.

 \begin{figure}[th]
\includegraphics[scale=0.4,bb=0 0 863 662]{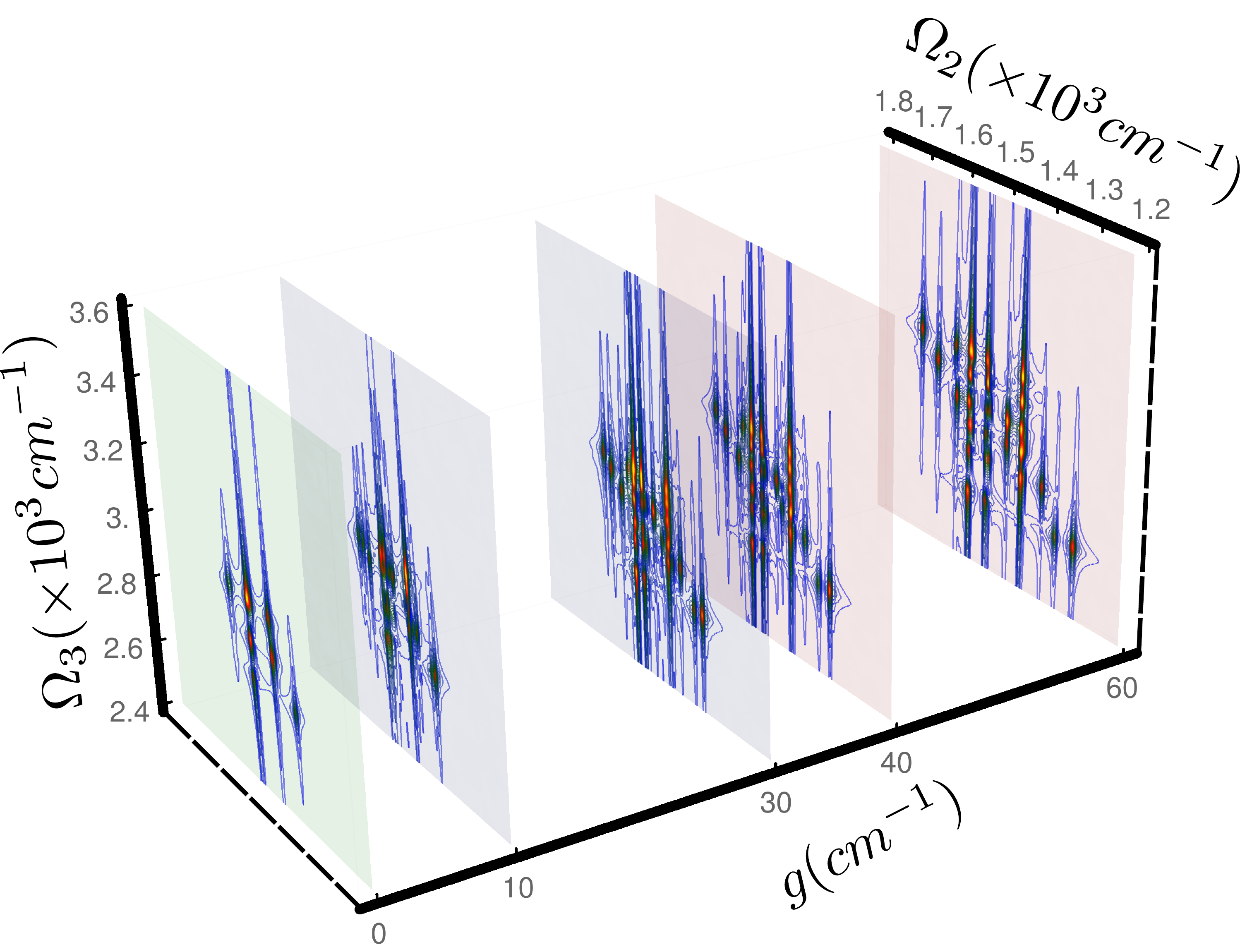}
\caption{Modulus of DQC signal for Amide-I+II vibrations in NMA with varying cavity coupling ($0 cm^{-1} \to 60 cm^{-1}$). Cavity field vector $\bm{e}_c$ is assumed to be mostly parallel to molecular vibrational transition dipole $\mu_m$. Amide-I vibrations ($\omega_1$) is resonant to cavity cutoff frequency $\omega_0=1625 cm^{-1}$ and $\omega_2=1540cm^{-1}$. The anharmonicies $\Delta_{11}=15cm^{-1}$ and $\Delta_{22}=11cm^{-1}$ (Table ~\ref{tb2}). \label{3d}}
\end{figure}

\begin{figure}[th]
\includegraphics[scale=0.28,bb=0 0 1724 1604]{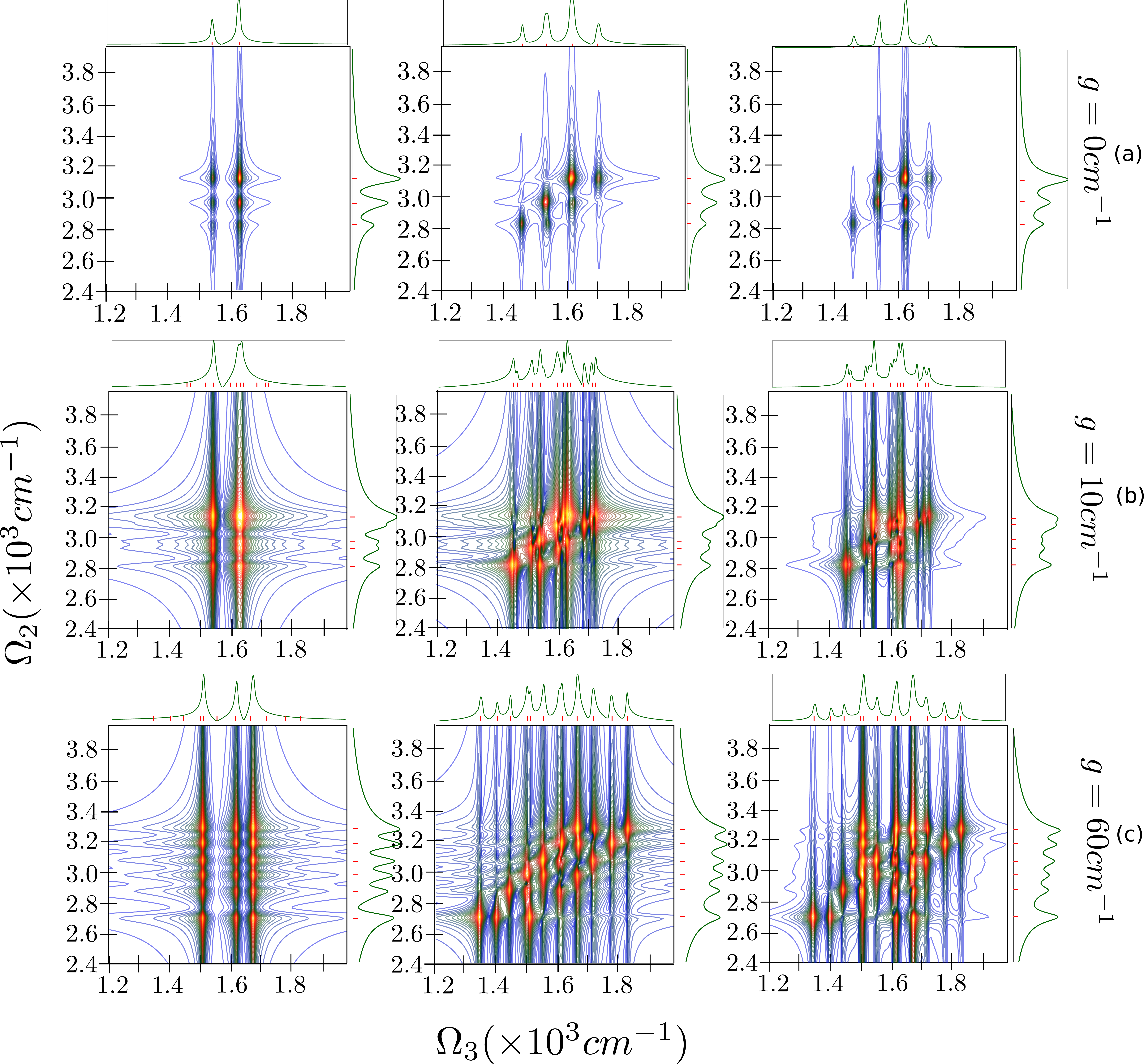}
\caption{DQC signals for Amide-I+II vibrations in NMA using Eq.~\ref{Sigplot} and ladder diagrams (Fig.~\ref{fig1a}e): $Columns:$ $(\text{left-}) |\mathcal{S}_i(\Omega_3,\Omega_2,0)| $ with resonance at $\Omega_{e_1g}$, $\Omega_{e_2g}$ and $\Omega_{e_3g}$ ; $ (\text{middle-})|\mathcal{S}_{ii}(\Omega_3,\Omega_2,0)| $ with resonance at $\Omega_{f_1e_3} \approx \Omega_{f_1e_2} $, $\Omega_{f_3e_4} \approx \Omega_{f_4e_2} $, $\Omega_{f_2e_3} \approx \Omega_{f_2e_2} $, $\Omega_{f_1e_1} $, $\Omega_{f_3e_3} \approx \Omega_{f_3e_2}=\Omega_{f_4e_1}=\Omega_{f_6e_3} $, $\Omega_{f_6e_2} $, $\Omega_{f_5e_3} $, $\Omega_{f_2e_1} $, $\Omega_{f_5e_2} $, $\Omega_{f_3e_1} $, $\Omega_{f_6e_1} $, $\Omega_{f_5e_1} $ and $(\text{right-})|\mathcal{S}(\Omega_3,\Omega_2,0)|$ with respective resonances and cavity couplings ($rows$) (a)$0 cm^{-1}$, (b)$10 cm^{-1}$, and (c) $60 cm^{-1}$ along $\Omega_3$, while peaks at $\Omega_{f_1g} $,$\Omega_{f_2g} $,$\Omega_{f_3g} $,$\Omega_{f_4g} $,$\Omega_{f_5g}$ and $\Omega_{f_6g}$ along $\Omega_2$. The linear projections along each axis is shown in green. For simplicity, cavity field vector $\bm{e}_c$ is assumed to be mostly parallel to molecular vibrational transition dipole $\mu_m$. Amide-I vibrations ($\omega_1$) is resonant to cavity cutoff frequency $\omega_0=1625 cm^{-1}$ and $\omega_2=1540cm^{-1}$. The anharmonicities are assumed $\Delta_{11}=15cm^{-1}$ and $\Delta_{22}=11cm^{-1}$ (Table ~\ref{tb2})  \label{sgcav2}}
\end{figure}

\textit{No cavity (Fig.~\ref{sgcav2})(a):} The coupled Amide-I+II vibrations of NMA is effectively a three level system with two singly excited states and three doubly excited states in absence of cavity coupling. The two singly excited states are resonant with $\Omega_{e_1g}$ and $\Omega{e_2g}$. The three doubly excited states are resonant with frequencies $\Omega_{f_1g}=2\Omega_{e_1g}-\Delta_{11} $, $\Omega_{f_2g}=\Omega_{e_1g}+\Omega_{e_2g} $ and $\Omega_{f_3g}=2\Omega_{e_2g}-\Delta_{22} $ (Sec. SII of supplementary material ~\cite{supp}, Table~\ref{tb2}). These peaks are observed along $\Omega_2$ axis. Whereas, we only observe four peaks along $\Omega_3$ axis in (Fig.~\ref{sgcav2}(a, $\text{right column}$)) because the tuples $(\Omega_{f_2e_1},\Omega_{f_1e_2}),\Omega_{e_1g})$ and $(\Omega_{f_3e_2},\Omega_{f_1e_1}),\Omega_{e_2g})$ cannot be resolved due to anharmonicities ($\Delta_{ii}$, Table ~\ref{tb2}).

\textit{Weak coupling regime with $\widetilde{g}=10cm^{-1}$ (Fig.~\ref{sgcav2})(b):} Under these conditions, we observe five peaks along $\Omega_2$ axis corresponding to $\Omega_{f_1g} $,$\Omega_{f_2g} $,$\Omega_{f_3g} $,$\Omega_{f_4g} $,$\Omega_{f_5g}=\Omega_{f_6g}$. Projections along $\Omega_3$ axis shows two peaks for $|\mathcal{S}_{i}(\Omega_3,\Omega_2,0)|$ (Fig.~\ref{sgcav2}(b,b$\text{left column}$)) resonant with frequencies $\Omega_{e_1g}$, $ \Omega_{e_2g} \approx \Omega_{e_3g}$. For $|\mathcal{S}_{ii}(\Omega_3,\Omega_2,0)|$ (Fig.~\ref{sgcav2}(b,b$\text{middle column}$)) we would ideally expect eighteen peaks corresponding to $\Omega_{f_1e_3} $, $\Omega_{f_1e_2} $, $\Omega_{f_3e_4} $, $\Omega_{f_4e_2} $, $\Omega_{f_2e_3} $, $\Omega_{f_2e_2} $, $\Omega_{f_1e_1} $, $\Omega_{f_3e_3} $, $\Omega_{f_3e_2}=\Omega_{f_4e_1}=\Omega_{f_6e_3} $, $\Omega_{f_6e_2} $, $\Omega_{f_5e_3} $, $\Omega_{f_2e_1} $, $\Omega_{f_5e_2} $, $\Omega_{f_3e_1} $, $\Omega_{f_6e_1} $, $\Omega_{f_5e_1} $ in energetically increasing order. However, only twelve peaks are observed due to overlapping caused by anharmonicities and coupling strength. Similar to Amide-I vibrations of NMA in cavity, we observe fewer peaks in  $|\mathcal{S}(\Omega_3,\Omega_2,0)|$ (Fig.~\ref{sgcav2}(b, $\text{right column}$)) than that of $|\mathcal{S}_{ii}(\Omega_3,\Omega_2,0)|$ due to destructive interferences between $|\mathcal{S}_{ii}(\Omega_3,\Omega_2,0)|$ and $|\mathcal{S}_{i}(\Omega_3,\Omega_2,0)|$.

\textit{Strong coupling regime $\widetilde{g}=50cm^{-1}$ (Fig.~\ref{sgcav2}(c)):} Increasing coupling resolves the six bipolariton resonances along $\Omega_2$ for all contributions to the DQC signal (Fig.~\ref{sgcav2}(c)). Along $\Omega_3$ axis, however, only $|\mathcal{S}_{i}(\Omega_3,\Omega_2,0)|$ is more resolved with three peaks corresponding to $\Omega_{e_1g} $, $\Omega_{e_2g} $ and $\Omega_{e_3g}$ with splitting $\sim \widetilde{g}_i$. Despite better peak resolution due to coupling strength dependent peak separations, resonances due to overlapping peaks cannot be fully resolved. The total signal (Fig.~\ref{sgcav2}($\text{right column}$)) only has 10 distinct peaks as a result of destructive interference of energies of $\Omega_{e_2g}$ and $\Omega_{e_3g}$ with some of the resonances from $|\mathcal{S}_{ii}(\Omega_3,\Omega_2,0)|$. 

Increasing coupling strengths provides us well resolved molecular vibrational bipolariton manifold structure (along $\Omega_2$ axis), however, it may not provide well resolved DQC signals along $\Omega_3$ axis requiring a careful tuning of $\widetilde{g}_{i}$.

\section{Conclusions}\label{CO}

The Multidimensional DQC signals shown in Figs.~\ref{schematic} - \ref{sgcav2} demonstrate how the ground state vibrational excitation manifolds are modified upon coupling to the cavity modes. The anharmonicities in the case of Amide-I of NMA coupled to cavity (for non-zero detuning  $\delta \neq 0$) are modified with cavity coupling strength, $\Delta_{ij}\rightarrow \widetilde{V}_{ij} = 16\Delta_{11} |\widetilde{g}_1^4/(\delta^2-16 \widetilde{g}_1^2)^2 |$ and this can be observed in the peak splitting. For zero detuning, the anharmonicitiy is independent of coupling strength, i.e., $\lim_{\delta \to 0}{\widetilde{V}_{ij}} = \Delta_{11}/32 $. However, in the case of Amide-I+II of NMA coupled to the infrared cavity anharmonicities in polariton basis varying nontrivially by, $\widetilde{V}_{ij}=1/2\Delta_{ij} |X_i|^2|X_j|^2$ where, $X_{i/j}$ depend on higher order of coupling strengths. This causes several peaks to overlap beyond certain cavity coupling and this is why not all expected peaks for $\Omega_{fe}$ can be spectrally resolved in Fig.~\ref{sgcav2}. However, we can use cavity coupling dependent anharmonicities for modifications of ground vibrational structures to mimic weakly interacting molecular vibrational multi-modes.   

Controlling and manipulating lifetime of single molecular vibrations by tuning the coupling strengths and time delays between ultrafast pulses (say, $t_1$) can reveal new energy transfer mechanisms. Varying cavity polarization ($\bm{\mu}_k\cdot \bm{e}_c$) and incident angle ($\theta$) in Eq.~\ref{ex-cavham1} using collinear experiments \cite{nardin2015multidimensional} could provide information regarding spatial confinements of several vibrational excitations and possibilities of molecular vibrational condensates like usual electronic polariton condensates in organic \cite{deng2010exciton,bittner2012,Zaster2015} and inorganic \cite{deng2010exciton,kasprzak2006bose,plumhof2014room} microcavities, if effective dense vibrational polaritons are obtained, which may be possible in larger macromolecules like J-aggregates. Time-varying cavity coupling strengths can be used to study dynamics of efficient cooling of molecular vibrational states for $n-polariton$ manifolds using higher dimensional spectroscopic techniques.

This work can be extended to control collective molecular vibrational excitations, which may be a useful tool in molecular cooling \cite{kowalewski2011cavity} allowing to perform ultracold experiments even at room-temperature. Furthermore, by varying the time delays one could be able to modify molecular vibrations via cavity and catch them in action in real time. We show that, by varying the cavity coupling strength $(\widetilde{g})$, it is possible to retrieve spectrally well-resolved molecular vibrational polariton(bipolariton) resonances which are otherwise difficult to resolve (see Fig.~\ref{schematic} - \ref{sgcav2} for example). Similar results with respect of the anharmonicities and peak redistributions due to cavity coupling can be achieved using other 2DIR techniques, e.g. $\bm{k_s}=-\bm k_1+\bm k_{2} + \bm k_{3}$ or $\bm{k_s}=\bm k_1-\bm k_{2} + \bm k_{3}$, however, the full power of DQC measurements presented here for single molecule in optical cavity can be seen more easily for macromolecules. For such larger systems, a different and much faster numerical algorithms utilizing techniques similar to Nonlinear Exciton Equations (NEE) \citep{abramavicius2009coherent} will be useful. Studying the dynamics of modified electronic ground states by applying methods developed in Ref.~\citep{mukamel1999principles} can also be done. In addition, incorporating electronic states can  similarly be achieved to modify vibrational excitations of electronically excited states.
\begin{acknowledgements}
We wish to thank Dr. Markus Kowalewski for valuable comments. The authors gratefully acknowledge the support of National Science Foundation (Grant No. CHE-1361516) and the Chemical Sciences, Geosciences and Biosciences Division, Office of Basic Energy Sciences, Office of Science, U.S. Department of Energy through award no. DE-FG02-4ER15571. The computational resources were provided by DOE.
\end{acknowledgements}


\begin{thebibliography}{41}
\expandafter\ifx\csname natexlab\endcsname\relax\def\natexlab#1{#1}\fi
\expandafter\ifx\csname bibnamefont\endcsname\relax
  \def\bibnamefont#1{#1}\fi
\expandafter\ifx\csname bibfnamefont\endcsname\relax
  \def\bibfnamefont#1{#1}\fi
\expandafter\ifx\csname citenamefont\endcsname\relax
  \def\citenamefont#1{#1}\fi
\expandafter\ifx\csname url\endcsname\relax
  \def\url#1{\texttt{#1}}\fi
\expandafter\ifx\csname urlprefix\endcsname\relax\def\urlprefix{URL }\fi
\providecommand{\bibinfo}[2]{#2}
\providecommand{\eprint}[2][]{\url{#2}}

\bibitem[{\citenamefont{Berman}(1994)}]{berman1994cavity}
\bibinfo{author}{\bibfnamefont{P.~R.} \bibnamefont{Berman}},
  \emph{\bibinfo{title}{{Cavity quantum electrodynamics}}}
  (\bibinfo{publisher}{Academic Press, Inc., Boston, MA (United States)},
  \bibinfo{year}{1994}).

\bibitem[{\citenamefont{Haroche and Raimond}(2006)}]{haroche2006exploring}
\bibinfo{author}{\bibfnamefont{S.}~\bibnamefont{Haroche}} \bibnamefont{and}
  \bibinfo{author}{\bibfnamefont{J.~M.} \bibnamefont{Raimond}},
  \emph{\bibinfo{title}{{Exploring the quantum}}} (\bibinfo{publisher}{Oxford
  Univ. Press}, \bibinfo{year}{2006}).

\bibitem[{\citenamefont{Coles et~al.}(2014{\natexlab{a}})\citenamefont{Coles,
  Yang, Wang, Grant, Taylor, Saikin, Aspuru-Guzik, Lidzey, Tang, and
  Smith}}]{coles2014strong}
\bibinfo{author}{\bibfnamefont{D.~M.} \bibnamefont{Coles}},
  \bibinfo{author}{\bibfnamefont{Y.}~\bibnamefont{Yang}},
  \bibinfo{author}{\bibfnamefont{Y.}~\bibnamefont{Wang}},
  \bibinfo{author}{\bibfnamefont{R.~T.} \bibnamefont{Grant}},
  \bibinfo{author}{\bibfnamefont{R.~A.} \bibnamefont{Taylor}},
  \bibinfo{author}{\bibfnamefont{S.~K.} \bibnamefont{Saikin}},
  \bibinfo{author}{\bibfnamefont{A.}~\bibnamefont{Aspuru-Guzik}},
  \bibinfo{author}{\bibfnamefont{D.~G.} \bibnamefont{Lidzey}},
  \bibinfo{author}{\bibfnamefont{J.~K.-H.} \bibnamefont{Tang}},
  \bibnamefont{and} \bibinfo{author}{\bibfnamefont{J.~M.} \bibnamefont{Smith}},
  \bibinfo{journal}{Nature communications} \textbf{\bibinfo{volume}{5}}
  (\bibinfo{year}{2014}{\natexlab{a}}).

\bibitem[{\citenamefont{Coles et~al.}(2014{\natexlab{b}})\citenamefont{Coles,
  Somaschi, Michetti, Clark, Lagoudakis, Savvidis, and
  Lidzey}}]{coles2014polariton}
\bibinfo{author}{\bibfnamefont{D.~M.} \bibnamefont{Coles}},
  \bibinfo{author}{\bibfnamefont{N.}~\bibnamefont{Somaschi}},
  \bibinfo{author}{\bibfnamefont{P.}~\bibnamefont{Michetti}},
  \bibinfo{author}{\bibfnamefont{C.}~\bibnamefont{Clark}},
  \bibinfo{author}{\bibfnamefont{P.~G.} \bibnamefont{Lagoudakis}},
  \bibinfo{author}{\bibfnamefont{P.~G.} \bibnamefont{Savvidis}},
  \bibnamefont{and} \bibinfo{author}{\bibfnamefont{D.~G.}
  \bibnamefont{Lidzey}}, \bibinfo{journal}{Nature materials}
  \textbf{\bibinfo{volume}{13}}, \bibinfo{pages}{712}
  (\bibinfo{year}{2014}{\natexlab{b}}).

\bibitem[{\citenamefont{Herrera et~al.}(2014)\citenamefont{Herrera, Peropadre,
  Pachon, Saikin, and Aspuru-Guzik}}]{herrera2014quantum}
\bibinfo{author}{\bibfnamefont{F.}~\bibnamefont{Herrera}},
  \bibinfo{author}{\bibfnamefont{B.}~\bibnamefont{Peropadre}},
  \bibinfo{author}{\bibfnamefont{L.~A.} \bibnamefont{Pachon}},
  \bibinfo{author}{\bibfnamefont{S.~K.} \bibnamefont{Saikin}},
  \bibnamefont{and}
  \bibinfo{author}{\bibfnamefont{A.}~\bibnamefont{Aspuru-Guzik}},
  \bibinfo{journal}{The Journal of Physical Chemistry Letters}
  \textbf{\bibinfo{volume}{5}}, \bibinfo{pages}{3708} (\bibinfo{year}{2014}).

\bibitem[{\citenamefont{Shalabney
  et~al.}(2015{\natexlab{a}})\citenamefont{Shalabney, George, Hutchison,
  Pupillo, Genet, and Ebbesen}}]{shalabney2015coherent}
\bibinfo{author}{\bibfnamefont{A.}~\bibnamefont{Shalabney}},
  \bibinfo{author}{\bibfnamefont{J.}~\bibnamefont{George}},
  \bibinfo{author}{\bibfnamefont{J.}~\bibnamefont{Hutchison}},
  \bibinfo{author}{\bibfnamefont{G.}~\bibnamefont{Pupillo}},
  \bibinfo{author}{\bibfnamefont{C.}~\bibnamefont{Genet}}, \bibnamefont{and}
  \bibinfo{author}{\bibfnamefont{T.~W.} \bibnamefont{Ebbesen}},
  \bibinfo{journal}{Nature communications} \textbf{\bibinfo{volume}{6}}
  (\bibinfo{year}{2015}{\natexlab{a}}).

\bibitem[{\citenamefont{del Pino et~al.}(2015)\citenamefont{del Pino, Feist,
  and Garcia-Vidal}}]{del2015quantum}
\bibinfo{author}{\bibfnamefont{J.}~\bibnamefont{del Pino}},
  \bibinfo{author}{\bibfnamefont{J.}~\bibnamefont{Feist}}, \bibnamefont{and}
  \bibinfo{author}{\bibfnamefont{F.~J.} \bibnamefont{Garcia-Vidal}},
  \bibinfo{journal}{New Journal of Physics} \textbf{\bibinfo{volume}{17}},
  \bibinfo{pages}{053040} (\bibinfo{year}{2015}).

\bibitem[{\citenamefont{Shalabney
  et~al.}(2015{\natexlab{b}})\citenamefont{Shalabney, George, Hiura, Hutchison,
  Genet, Hellwig, and Ebbesen}}]{shalabney2015enhanced}
\bibinfo{author}{\bibfnamefont{A.}~\bibnamefont{Shalabney}},
  \bibinfo{author}{\bibfnamefont{J.}~\bibnamefont{George}},
  \bibinfo{author}{\bibfnamefont{H.}~\bibnamefont{Hiura}},
  \bibinfo{author}{\bibfnamefont{J.~A.} \bibnamefont{Hutchison}},
  \bibinfo{author}{\bibfnamefont{C.}~\bibnamefont{Genet}},
  \bibinfo{author}{\bibfnamefont{P.}~\bibnamefont{Hellwig}}, \bibnamefont{and}
  \bibinfo{author}{\bibfnamefont{T.~W.} \bibnamefont{Ebbesen}},
  \bibinfo{journal}{Angewandte Chemie International Edition}
  \textbf{\bibinfo{volume}{54}}, \bibinfo{pages}{7971}
  (\bibinfo{year}{2015}{\natexlab{b}}).

\bibitem[{\citenamefont{Simpkins et~al.}(2015)\citenamefont{Simpkins, Fears,
  Dressick, Spann, Dunkelberger, and Owrutsky}}]{simpkins2015spanning}
\bibinfo{author}{\bibfnamefont{B.}~\bibnamefont{Simpkins}},
  \bibinfo{author}{\bibfnamefont{K.~P.} \bibnamefont{Fears}},
  \bibinfo{author}{\bibfnamefont{W.~J.} \bibnamefont{Dressick}},
  \bibinfo{author}{\bibfnamefont{B.~T.} \bibnamefont{Spann}},
  \bibinfo{author}{\bibfnamefont{A.~D.} \bibnamefont{Dunkelberger}},
  \bibnamefont{and} \bibinfo{author}{\bibfnamefont{J.~C.}
  \bibnamefont{Owrutsky}}, \bibinfo{journal}{ACS Photonics}
  (\bibinfo{year}{2015}).

\bibitem[{\citenamefont{Autry et~al.}(2015)\citenamefont{Autry, Nardin, Bajoni,
  Lema{\^\i}tre, Bouchoule, Bloch, and Cundiff}}]{autry2015measurement}
\bibinfo{author}{\bibfnamefont{T.}~\bibnamefont{Autry}},
  \bibinfo{author}{\bibfnamefont{G.}~\bibnamefont{Nardin}},
  \bibinfo{author}{\bibfnamefont{D.}~\bibnamefont{Bajoni}},
  \bibinfo{author}{\bibfnamefont{A.}~\bibnamefont{Lema{\^\i}tre}},
  \bibinfo{author}{\bibfnamefont{S.}~\bibnamefont{Bouchoule}},
  \bibinfo{author}{\bibfnamefont{J.}~\bibnamefont{Bloch}}, \bibnamefont{and}
  \bibinfo{author}{\bibfnamefont{S.}~\bibnamefont{Cundiff}}, in
  \emph{\bibinfo{booktitle}{{CLEO: QELS\_Fundamental Science}}}
  (\bibinfo{organization}{Optical Society of America}, \bibinfo{year}{2015}),
  pp. \bibinfo{pages}{FW4B--7}.

\bibitem[{\citenamefont{Sun et~al.}(2015)\citenamefont{Sun, Yoon, Steger, Liu,
  Pfeiffer, West, Snoke, and Nelson}}]{sun2015polaritons}
\bibinfo{author}{\bibfnamefont{Y.}~\bibnamefont{Sun}},
  \bibinfo{author}{\bibfnamefont{Y.}~\bibnamefont{Yoon}},
  \bibinfo{author}{\bibfnamefont{M.}~\bibnamefont{Steger}},
  \bibinfo{author}{\bibfnamefont{G.}~\bibnamefont{Liu}},
  \bibinfo{author}{\bibfnamefont{L.~N.} \bibnamefont{Pfeiffer}},
  \bibinfo{author}{\bibfnamefont{K.}~\bibnamefont{West}},
  \bibinfo{author}{\bibfnamefont{D.~W.} \bibnamefont{Snoke}}, \bibnamefont{and}
  \bibinfo{author}{\bibfnamefont{K.~A.} \bibnamefont{Nelson}},
  \bibinfo{journal}{arXiv preprint arXiv:1508.06698}  (\bibinfo{year}{2015}).

\bibitem[{\citenamefont{Wilmer et~al.}(2015)\citenamefont{Wilmer, Passmann,
  Gehl, Khitrova, and Bristow}}]{wilmer2015multidimensional}
\bibinfo{author}{\bibfnamefont{B.~L.} \bibnamefont{Wilmer}},
  \bibinfo{author}{\bibfnamefont{F.}~\bibnamefont{Passmann}},
  \bibinfo{author}{\bibfnamefont{M.}~\bibnamefont{Gehl}},
  \bibinfo{author}{\bibfnamefont{G.}~\bibnamefont{Khitrova}}, \bibnamefont{and}
  \bibinfo{author}{\bibfnamefont{A.~D.} \bibnamefont{Bristow}},
  \bibinfo{journal}{Physical Review B} \textbf{\bibinfo{volume}{91}},
  \bibinfo{pages}{201304} (\bibinfo{year}{2015}).

\bibitem[{\citenamefont{Savona et~al.}(1995)\citenamefont{Savona, Andreani,
  Schwendimann, and Quattropani}}]{savona1995quantum}
\bibinfo{author}{\bibfnamefont{V.}~\bibnamefont{Savona}},
  \bibinfo{author}{\bibfnamefont{L.}~\bibnamefont{Andreani}},
  \bibinfo{author}{\bibfnamefont{P.}~\bibnamefont{Schwendimann}},
  \bibnamefont{and}
  \bibinfo{author}{\bibfnamefont{A.}~\bibnamefont{Quattropani}},
  \bibinfo{journal}{Solid State Communications} \textbf{\bibinfo{volume}{93}},
  \bibinfo{pages}{733} (\bibinfo{year}{1995}).

\bibitem[{\citenamefont{Feist and Garcia-Vidal}(2015)}]{feist2015extraordinary}
\bibinfo{author}{\bibfnamefont{J.}~\bibnamefont{Feist}} \bibnamefont{and}
  \bibinfo{author}{\bibfnamefont{F.~J.} \bibnamefont{Garcia-Vidal}},
  \bibinfo{journal}{Physical Review Letters} \textbf{\bibinfo{volume}{114}},
  \bibinfo{pages}{196402} (\bibinfo{year}{2015}).

\bibitem[{\citenamefont{Faez et~al.}(2014)\citenamefont{Faez, T{\"u}rschmann,
  Haakh, G{\"o}tzinger, and Sandoghdar}}]{faez2014coherent}
\bibinfo{author}{\bibfnamefont{S.}~\bibnamefont{Faez}},
  \bibinfo{author}{\bibfnamefont{P.}~\bibnamefont{T{\"u}rschmann}},
  \bibinfo{author}{\bibfnamefont{H.~R.} \bibnamefont{Haakh}},
  \bibinfo{author}{\bibfnamefont{S.}~\bibnamefont{G{\"o}tzinger}},
  \bibnamefont{and}
  \bibinfo{author}{\bibfnamefont{V.}~\bibnamefont{Sandoghdar}},
  \bibinfo{journal}{Physical review letters} \textbf{\bibinfo{volume}{113}},
  \bibinfo{pages}{213601} (\bibinfo{year}{2014}).
  
  

\bibitem[{\citenamefont{Al{\`u} et~al.}(2007)\citenamefont{Al{\`u},
  Silveirinha, Salandrino, and Engheta}}]{alu2007epsilon}
\bibinfo{author}{\bibfnamefont{A.}~\bibnamefont{Al{\`u}}},
  \bibinfo{author}{\bibfnamefont{M.~G.} \bibnamefont{Silveirinha}},
  \bibinfo{author}{\bibfnamefont{A.}~\bibnamefont{Salandrino}},
  \bibnamefont{and} \bibinfo{author}{\bibfnamefont{N.}~\bibnamefont{Engheta}},
  \bibinfo{journal}{Physical Review B} \textbf{\bibinfo{volume}{75}},
  \bibinfo{pages}{155410} (\bibinfo{year}{2007}).

\bibitem[{\citenamefont{Fleury and Al{\`u}}(2013)}]{fleury2013enhanced}
\bibinfo{author}{\bibfnamefont{R.}~\bibnamefont{Fleury}} \bibnamefont{and}
  \bibinfo{author}{\bibfnamefont{A.}~\bibnamefont{Al{\`u}}},
  \bibinfo{journal}{Physical Review B} \textbf{\bibinfo{volume}{87}},
  \bibinfo{pages}{201101} (\bibinfo{year}{2013}).

\bibitem[{\citenamefont{Kelkar et~al.}(2015)\citenamefont{Kelkar, Wang,
  Hoffmann, Christiansen, G{\"o}tzinger, and Sandoghdar}}]{kelkar2015sub}
\bibinfo{author}{\bibfnamefont{H.}~\bibnamefont{Kelkar}},
  \bibinfo{author}{\bibfnamefont{D.}~\bibnamefont{Wang}},
  \bibinfo{author}{\bibfnamefont{B.}~\bibnamefont{Hoffmann}},
  \bibinfo{author}{\bibfnamefont{S.}~\bibnamefont{Christiansen}},
  \bibinfo{author}{\bibfnamefont{S.}~\bibnamefont{G{\"o}tzinger}},
  \bibnamefont{and}
  \bibinfo{author}{\bibfnamefont{V.}~\bibnamefont{Sandoghdar}}, in
  \emph{\bibinfo{booktitle}{{European Quantum Electronics Conference}}}
  (\bibinfo{organization}{Optical Society of America}, \bibinfo{year}{2015}),
  p. \bibinfo{pages}{EG\_6\_1}.

\bibitem[{\citenamefont{Benz et~al.}(2015)\citenamefont{Benz, Campione, Klem,
  Sinclair, and Brener}}]{benz2015control}
\bibinfo{author}{\bibfnamefont{A.}~\bibnamefont{Benz}},
  \bibinfo{author}{\bibfnamefont{S.}~\bibnamefont{Campione}},
  \bibinfo{author}{\bibfnamefont{J.~F.} \bibnamefont{Klem}},
  \bibinfo{author}{\bibfnamefont{M.~B.} \bibnamefont{Sinclair}},
  \bibnamefont{and} \bibinfo{author}{\bibfnamefont{I.}~\bibnamefont{Brener}},
  \bibinfo{journal}{Nano letters} \textbf{\bibinfo{volume}{15}},
  \bibinfo{pages}{1959} (\bibinfo{year}{2015}).

\bibitem[{\citenamefont{Lai et~al.}(2013)\citenamefont{Lai, Preketes, Mukamel,
  and Wang}}]{lai2013monitoring}
\bibinfo{author}{\bibfnamefont{Z.}~\bibnamefont{Lai}},
  \bibinfo{author}{\bibfnamefont{N.~K.} \bibnamefont{Preketes}},
  \bibinfo{author}{\bibfnamefont{S.}~\bibnamefont{Mukamel}}, \bibnamefont{and}
  \bibinfo{author}{\bibfnamefont{J.}~\bibnamefont{Wang}}, \bibinfo{journal}{The
  Journal of Physical Chemistry B} \textbf{\bibinfo{volume}{117}},
  \bibinfo{pages}{4661} (\bibinfo{year}{2013}).

\bibitem[{\citenamefont{Abramavicius et~al.}(2008)\citenamefont{Abramavicius,
  Voronine, and Mukamel}}]{abramavicius2008double}
\bibinfo{author}{\bibfnamefont{D.}~\bibnamefont{Abramavicius}},
  \bibinfo{author}{\bibfnamefont{D.~V.} \bibnamefont{Voronine}},
  \bibnamefont{and} \bibinfo{author}{\bibfnamefont{S.}~\bibnamefont{Mukamel}},
  \bibinfo{journal}{Proceedings of the National Academy of Sciences}
  \textbf{\bibinfo{volume}{105}}, \bibinfo{pages}{8525} (\bibinfo{year}{2008}).

\bibitem[{\citenamefont{Venkatramani and
  Mukamel}(2002)}]{venkatramani2002correlated}
\bibinfo{author}{\bibfnamefont{R.}~\bibnamefont{Venkatramani}}
  \bibnamefont{and} \bibinfo{author}{\bibfnamefont{S.}~\bibnamefont{Mukamel}},
  \bibinfo{journal}{The Journal of chemical physics}
  \textbf{\bibinfo{volume}{117}}, \bibinfo{pages}{11089}
  (\bibinfo{year}{2002}).

\bibitem[{\citenamefont{Piryatinski et~al.}(2001)\citenamefont{Piryatinski,
  Chernyak, and Mukamel}}]{piryatinski2001two}
\bibinfo{author}{\bibfnamefont{A.}~\bibnamefont{Piryatinski}},
  \bibinfo{author}{\bibfnamefont{V.}~\bibnamefont{Chernyak}}, \bibnamefont{and}
  \bibinfo{author}{\bibfnamefont{S.}~\bibnamefont{Mukamel}},
  \bibinfo{journal}{Chemical Physics} \textbf{\bibinfo{volume}{266}},
  \bibinfo{pages}{311} (\bibinfo{year}{2001}).

\bibitem[{\citenamefont{Greve et~al.}(2013)\citenamefont{Greve, Preketes,
  Fidder, Costard, Koeppe, Heisler, Mukamel, Temps, Nibbering, and
  Elsaesser}}]{greve2013n}
\bibinfo{author}{\bibfnamefont{C.}~\bibnamefont{Greve}},
  \bibinfo{author}{\bibfnamefont{N.~K.} \bibnamefont{Preketes}},
  \bibinfo{author}{\bibfnamefont{H.}~\bibnamefont{Fidder}},
  \bibinfo{author}{\bibfnamefont{R.}~\bibnamefont{Costard}},
  \bibinfo{author}{\bibfnamefont{B.}~\bibnamefont{Koeppe}},
  \bibinfo{author}{\bibfnamefont{I.~A.} \bibnamefont{Heisler}},
  \bibinfo{author}{\bibfnamefont{S.}~\bibnamefont{Mukamel}},
  \bibinfo{author}{\bibfnamefont{F.}~\bibnamefont{Temps}},
  \bibinfo{author}{\bibfnamefont{E.~T.} \bibnamefont{Nibbering}},
  \bibnamefont{and}
  \bibinfo{author}{\bibfnamefont{T.}~\bibnamefont{Elsaesser}},
  \bibinfo{journal}{The Journal of Physical Chemistry A}
  \textbf{\bibinfo{volume}{117}}, \bibinfo{pages}{594} (\bibinfo{year}{2013}).

\bibitem[{\citenamefont{Cwik et~al.}(2015)\citenamefont{Cwik, Kirton, {De
  Liberato}, and Keeling}}]{cwik2015self}
\bibinfo{author}{\bibfnamefont{J.~A.} \bibnamefont{Cwik}},
  \bibinfo{author}{\bibfnamefont{P.}~\bibnamefont{Kirton}},
  \bibinfo{author}{\bibfnamefont{S.}~\bibnamefont{{De Liberato}}},
  \bibnamefont{and} \bibinfo{author}{\bibfnamefont{J.}~\bibnamefont{Keeling}},
  \bibinfo{journal}{arXiv preprint arXiv:1506.08974}  (\bibinfo{year}{2015}).

\bibitem[{\citenamefont{Mukamel}(1999)}]{mukamel1999principles}
\bibinfo{author}{\bibfnamefont{S.}~\bibnamefont{Mukamel}},
  \emph{\bibinfo{title}{{Principles of nonlinear optical spectroscopy}}},
  \bibinfo{number}{6} (\bibinfo{publisher}{Oxford University Press},
  \bibinfo{year}{1999}).

\bibitem[{\citenamefont{Nardin}(2015)}]{nardin2015multidimensional}
\bibinfo{author}{\bibfnamefont{G.}~\bibnamefont{Nardin}},
  \bibinfo{journal}{Semiconductor Science and Technology}
  \textbf{\bibinfo{volume}{31}}, \bibinfo{pages}{023001}
  (\bibinfo{year}{2015}).

\bibitem[{\citenamefont{Takemura et~al.}(2015)\citenamefont{Takemura, Trebaol,
  Anderson, Kohnle, L{\'e}ger, Oberli, Portella-Oberli, and
  Deveaud}}]{takemura2015two}
\bibinfo{author}{\bibfnamefont{N.}~\bibnamefont{Takemura}},
  \bibinfo{author}{\bibfnamefont{S.}~\bibnamefont{Trebaol}},
  \bibinfo{author}{\bibfnamefont{M.}~\bibnamefont{Anderson}},
  \bibinfo{author}{\bibfnamefont{V.}~\bibnamefont{Kohnle}},
  \bibinfo{author}{\bibfnamefont{Y.}~\bibnamefont{L{\'e}ger}},
  \bibinfo{author}{\bibfnamefont{D.}~\bibnamefont{Oberli}},
  \bibinfo{author}{\bibfnamefont{M.~T.} \bibnamefont{Portella-Oberli}},
  \bibnamefont{and} \bibinfo{author}{\bibfnamefont{B.}~\bibnamefont{Deveaud}},
  \bibinfo{journal}{Physical Review B} \textbf{\bibinfo{volume}{92}},
  \bibinfo{pages}{125415} (\bibinfo{year}{2015}).

\bibitem[{\citenamefont{Oka and Ishihara}(2008{\natexlab{a}})}]{oka2008real}
\bibinfo{author}{\bibfnamefont{H.}~\bibnamefont{Oka}} \bibnamefont{and}
  \bibinfo{author}{\bibfnamefont{H.}~\bibnamefont{Ishihara}},
  \bibinfo{journal}{Physical Review B} \textbf{\bibinfo{volume}{78}},
  \bibinfo{pages}{195314} (\bibinfo{year}{2008}{\natexlab{a}}).

\bibitem[{\citenamefont{Oka and Ishihara}(2008{\natexlab{b}})}]{oka2008highly}
\bibinfo{author}{\bibfnamefont{H.}~\bibnamefont{Oka}} \bibnamefont{and}
  \bibinfo{author}{\bibfnamefont{H.}~\bibnamefont{Ishihara}},
  \bibinfo{journal}{Physical review letters} \textbf{\bibinfo{volume}{100}},
  \bibinfo{pages}{170505} (\bibinfo{year}{2008}{\natexlab{b}}).

\bibitem[{\citenamefont{Chernyak et~al.}(1998)\citenamefont{Chernyak, Zhang,
  and Mukamel}}]{chernyak1998multidimensional}
\bibinfo{author}{\bibfnamefont{V.}~\bibnamefont{Chernyak}},
  \bibinfo{author}{\bibfnamefont{W.~M.} \bibnamefont{Zhang}}, \bibnamefont{and}
  \bibinfo{author}{\bibfnamefont{S.}~\bibnamefont{Mukamel}},
  \bibinfo{journal}{The Journal of chemical physics}
  \textbf{\bibinfo{volume}{109}}, \bibinfo{pages}{9587} (\bibinfo{year}{1998}).

\bibitem[{\citenamefont{Abramavicius et~al.}(2009)\citenamefont{Abramavicius,
  Palmieri, Voronine, Sanda, and Mukamel}}]{abramavicius2009coherent}
\bibinfo{author}{\bibfnamefont{D.}~\bibnamefont{Abramavicius}},
  \bibinfo{author}{\bibfnamefont{B.}~\bibnamefont{Palmieri}},
  \bibinfo{author}{\bibfnamefont{D.~V.} \bibnamefont{Voronine}},
  \bibinfo{author}{\bibfnamefont{F.}~\bibnamefont{Sanda}}, \bibnamefont{and}
  \bibinfo{author}{\bibfnamefont{S.}~\bibnamefont{Mukamel}},
  \bibinfo{journal}{Chemical reviews} \textbf{\bibinfo{volume}{109}},
  \bibinfo{pages}{2350} (\bibinfo{year}{2009}).

\bibitem[{\citenamefont{Roslyak et~al.}(2010)\citenamefont{Roslyak, Gumbs, and
  Mukamel}}]{roslyak2010two}
\bibinfo{author}{\bibfnamefont{O.}~\bibnamefont{Roslyak}},
  \bibinfo{author}{\bibfnamefont{G.}~\bibnamefont{Gumbs}}, \bibnamefont{and}
  \bibinfo{author}{\bibfnamefont{S.}~\bibnamefont{Mukamel}},
  \bibinfo{journal}{Nano letters} \textbf{\bibinfo{volume}{10}},
  \bibinfo{pages}{4253} (\bibinfo{year}{2010}).


  
\bibitem[]{supp} \text{See supplemental material at [URL will be inserted by AIP] } \text{for total Hamiltonian of system in polariton basis and} \text{respective singly and doubly excited manifold (Eqs. S1-S8).} \text{Figures S1-S10 show dispersive and absorptive parts of the} \text{DQC signals shown in maintext. Figure S11-S12 shows} \text{respective peak splittings.}

\bibitem[{\citenamefont{Hayashi and Mukamel}(2008)}]{hayashi2008two}
\bibinfo{author}{\bibfnamefont{T.}~\bibnamefont{Hayashi}} \bibnamefont{and}
  \bibinfo{author}{\bibfnamefont{S.}~\bibnamefont{Mukamel}},
  \bibinfo{journal}{Journal of molecular liquids}
  \textbf{\bibinfo{volume}{141}}, \bibinfo{pages}{149} (\bibinfo{year}{2008}).

\bibitem[{\citenamefont{Rubtsov et~al.}(2003)\citenamefont{Rubtsov, Wang, and
  Hochstrasser}}]{rubtsov2003vibrational}
\bibinfo{author}{\bibfnamefont{I.~V.} \bibnamefont{Rubtsov}},
  \bibinfo{author}{\bibfnamefont{J.}~\bibnamefont{Wang}}, \bibnamefont{and}
  \bibinfo{author}{\bibfnamefont{R.~M.} \bibnamefont{Hochstrasser}},
  \bibinfo{journal}{The Journal of Physical Chemistry A}
  \textbf{\bibinfo{volume}{107}}, \bibinfo{pages}{3384} (\bibinfo{year}{2003}).

\bibitem[{\citenamefont{Barth}(2007)}]{barth2007infrared}
\bibinfo{author}{\bibfnamefont{A.}~\bibnamefont{Barth}},
  \bibinfo{journal}{Biochimica et Biophysica Acta (BBA)-Bioenergetics}
  \textbf{\bibinfo{volume}{1767}}, \bibinfo{pages}{1073}
  (\bibinfo{year}{2007}).

\bibitem[{\citenamefont{DeFlores et~al.}(2006)\citenamefont{DeFlores, Ganim,
  Ackley, Chung, and Tokmakoff}}]{deflores2006anharmonic}
\bibinfo{author}{\bibfnamefont{L.~P.} \bibnamefont{DeFlores}},
  \bibinfo{author}{\bibfnamefont{Z.}~\bibnamefont{Ganim}},
  \bibinfo{author}{\bibfnamefont{S.~F.} \bibnamefont{Ackley}},
  \bibinfo{author}{\bibfnamefont{H.~S.} \bibnamefont{Chung}}, \bibnamefont{and}
  \bibinfo{author}{\bibfnamefont{A.}~\bibnamefont{Tokmakoff}},
  \bibinfo{journal}{The Journal of Physical Chemistry B}
  \textbf{\bibinfo{volume}{110}}, \bibinfo{pages}{18973}
  (\bibinfo{year}{2006}).


\bibitem[{\citenamefont{Deng et~al.}(2010)\citenamefont{Deng, Haug, and
  Yamamoto}}]{deng2010exciton}
\bibinfo{author}{\bibfnamefont{H.}~\bibnamefont{Deng}},
  \bibinfo{author}{\bibfnamefont{H.}~\bibnamefont{Haug}}, \bibnamefont{and}
  \bibinfo{author}{\bibfnamefont{Y.}~\bibnamefont{Yamamoto}},
  \bibinfo{journal}{Reviews of modern physics} \textbf{\bibinfo{volume}{82}},
  \bibinfo{pages}{1489} (\bibinfo{year}{2010}).

 \bibitem[{\citenamefont{Bittner et~al.}(2012)\citenamefont{Bittner, Zaster, and
  Silva}}]{bittner2012}
\bibinfo{author}{\bibfnamefont{E.R.}~\bibnamefont{Bittner}},
  \bibinfo{author}{\bibfnamefont{S.}~\bibnamefont{Zaster}}, \bibnamefont{and}
  \bibinfo{author}{\bibfnamefont{C.}~\bibnamefont{Silva}},
  \bibinfo{journal}{Phys. Chem. Chem. Phys,} \textbf{\bibinfo{volume}{14}},
  \bibinfo{pages}{3226-3233} (\bibinfo{year}{2012}).

  \bibitem[{\citenamefont{Zaster et~al.}(2010)\citenamefont{Zaster, and
  Bittner}}]{Zaster2015}
\bibinfo{author}{\bibfnamefont{S.}~\bibnamefont{Zaster}},
  \bibnamefont{and}   \bibinfo{author}{\bibfnamefont{E.~R.}~\bibnamefont{Bittner}},
  \bibinfo{journal}{International Journal of Modern Physics B } \textbf{\bibinfo{volume}{29}},
  \bibinfo{pages}{1550157} (\bibinfo{year}{2015}).

\bibitem[{\citenamefont{Kasprzak et~al.}(2006)\citenamefont{Kasprzak, Richard,
  Kundermann, Baas, Jeambrun, Keeling, Marchetti, Szyma{\'n}ska, Andre, Staehli
  et~al.}}]{kasprzak2006bose}
  \bibinfo{author}{\bibfnamefont{J.}~\bibnamefont{Kasprzak}},
  \bibinfo{author}{\bibfnamefont{M.}~\bibnamefont{Richard}},
  \bibinfo{author}{\bibfnamefont{S.}~\bibnamefont{Kundermann}},
  \bibinfo{author}{\bibfnamefont{A.}~\bibnamefont{Baas}},
  \bibinfo{author}{\bibfnamefont{P.}~\bibnamefont{Jeambrun}},
  \bibinfo{author}{\bibfnamefont{J.}~\bibnamefont{Keeling}},
  \bibinfo{author}{\bibfnamefont{F.}~\bibnamefont{Marchetti}},
  \bibinfo{author}{\bibfnamefont{M.}~\bibnamefont{Szyma{\'n}ska}},
  \bibinfo{author}{\bibfnamefont{R.}~\bibnamefont{Andre}},
  \bibinfo{author}{\bibfnamefont{J.}~\bibnamefont{Staehli}},
  \bibnamefont{et~al.}, \bibinfo{journal}{Nature}
  \textbf{\bibinfo{volume}{443}}, \bibinfo{pages}{409} (\bibinfo{year}{2006}).

\bibitem[{\citenamefont{Plumhof et~al.}(2014)\citenamefont{Plumhof,
  St{\"o}ferle, Mai, Scherf, and Mahrt}}]{plumhof2014room}
\bibinfo{author}{\bibfnamefont{J.~D.} \bibnamefont{Plumhof}},
  \bibinfo{author}{\bibfnamefont{T.}~\bibnamefont{St{\"o}ferle}},
  \bibinfo{author}{\bibfnamefont{L.}~\bibnamefont{Mai}},
  \bibinfo{author}{\bibfnamefont{U.}~\bibnamefont{Scherf}}, \bibnamefont{and}
  \bibinfo{author}{\bibfnamefont{R.~F.} \bibnamefont{Mahrt}},
  \bibinfo{journal}{Nature materials} \textbf{\bibinfo{volume}{13}},
  \bibinfo{pages}{247} (\bibinfo{year}{2014}).

\bibitem[{\citenamefont{Kowalewski et~al.}(2011)\citenamefont{Kowalewski,
  Morigi, Pinkse, and de~Vivie-Riedle}}]{kowalewski2011cavity}
\bibinfo{author}{\bibfnamefont{M.}~\bibnamefont{Kowalewski}},
  \bibinfo{author}{\bibfnamefont{G.}~\bibnamefont{Morigi}},
  \bibinfo{author}{\bibfnamefont{P.~W.} \bibnamefont{Pinkse}},
  \bibnamefont{and}
  \bibinfo{author}{\bibfnamefont{R.}~\bibnamefont{de~Vivie-Riedle}},
  \bibinfo{journal}{Physical Review A} \textbf{\bibinfo{volume}{84}},
  \bibinfo{pages}{033408} (\bibinfo{year}{2011}).

\end{thebibliography}

\end{document}